\documentclass[conference,a4paper]{IEEEtran}
\usepackage{xcolor}
\usepackage{balance}
\usepackage{amsmath} 
\usepackage{makecell}
\usepackage{multirow}

\ifCLASSINFOpdf
  \usepackage[pdftex]{graphicx}
\else
  \usepackage[dvips]{graphicx}
\fi
\ifCLASSOPTIONcompsoc
 \usepackage[caption=false,font=normalsize,labelfont=sf,textfont=sf]{subfig}
\else
 \usepackage[caption=false,font=footnotesize]{subfig}
\fi
\begin{document}
%
\title{Terahertz Signal Coverage Enhancement in Hall Scenarios Based on Single-Hop and Dual-Hop Reconfigurable Intelligent Surfaces}

\author{\IEEEauthorblockN{
Ben Chen\IEEEauthorrefmark{1}\IEEEauthorrefmark{2},   
Zhangdui Zhong\IEEEauthorrefmark{2}\IEEEauthorrefmark{1},
Ke Guan\IEEEauthorrefmark{1}\IEEEauthorrefmark{2},   
Danping He\IEEEauthorrefmark{2}\IEEEauthorrefmark{1},    
Yiran Wang\IEEEauthorrefmark{1}\IEEEauthorrefmark{2},      
Jianwen Ding\IEEEauthorrefmark{1}\IEEEauthorrefmark{2}\IEEEauthorrefmark{3},
Qi Luo\IEEEauthorrefmark{4}    
}                                     
\IEEEauthorblockA{\IEEEauthorrefmark{1}
State Key Laboratory of Advanced Rail Autonomous Operation, Beijing Jiaotong University, Beijing, China}
\IEEEauthorblockA{\IEEEauthorrefmark{2}
School of Electronic and Information Engineering, Beijing Jiaotong University, Beijing, China}
\IEEEauthorblockA{\IEEEauthorrefmark{3}
Key Laboratory of Railway Industry of Broadband Mobile Information Communications, Beijing Jiaotong University, China}
\IEEEauthorblockA{\IEEEauthorrefmark{4}
 School of Physics, Engineering and Computer Science, University of Hertfordshire, Hatfield, United Kingdom}
  \IEEEauthorblockA{ Email: kguan@bjtu.edu.cn }
}



\maketitle

\begin{abstract}
Terahertz (THz) communication offers ultra-high data rates and has emerged as a promising technology for future wireless networks. However, the inherently high free-space path loss of THz waves significantly limits the coverage range of THz communication systems. Therefore, extending the effective coverage area is a key challenge for the practical deployment of THz networks. Reconfigurable intelligent surfaces (RIS), which can dynamically manipulate electromagnetic wave propagation, provide a solution to enhance THz coverage. To investigate multi-RIS deployment scenarios, this work integrates an antenna array–based RIS model into the ray-tracing simulation platform. Using an indoor hall as a representative case study, the enhancement effects of single-hop and dual-hop RIS configurations on indoor signal coverage are evaluated under various deployment schemes. The developed framework offers valuable insights and design references for optimizing RIS-assisted indoor THz communication and coverage~estimation.

\end{abstract}

\vskip0.5\baselineskip
\begin{IEEEkeywords}
 Indoor hall scenario, ray-tracing,  Reconfigurable Intelligent Surfaces, terahertz.
\end{IEEEkeywords}

%

\section{Introduction}

The terahertz (THz) band offers a vast and underutilized spectrum resource, with its expansive bandwidth characteristics enabling ultra-high data rates and large communication capacity \cite{Terahertz}\cite{Terahertz1}. However, the high frequency leads to significant path loss, resulting in severe attenuation of THz communication signals and restricting the effective coverage area \cite{PL1}\cite{PL2}. This challenge limits the wide deployment of THz~communication~technology.

In recent years, significant advancements have been made in the development of Reconfigurable Intelligent Surfaces (RIS). With its ability to control the propagation of electromagnetic waves, RIS has the potential to substantially enhance communication channel quality by reshaping the electromagnetic environment \cite{RIS}. Particularly in the THz band, the deployment of RIS has shown promise in overcoming path loss challenges, providing a new solution to improve signal coverage \cite{THZRIS}.

Existing research has demonstrated significant advancements in the use of RIS for THz communication. The authors of \cite{TUBSRISmeasure} conducted channel measurements with passive RIS at 300 GHz, and investigated the effects of the distance between the RIS and the receiver, as well as the angle of the receiver relative to the RIS. They also showed that the bistatic radar equation is suitable for path gain prediction in this scenario. The authors of \cite{TRISRRIS} presented measurements on both reflective and transmissive RIS in the THz band, revealing that the narrow half-power beamwidths (HPBW) of the reflected or transmitted beams make them sensitive to slight alignment deviations, which can affect the channel~gain.

Ray-tracing (RT) technology, as a deterministic channel modeling method, effectively simulates the electromagnetic propagation in real-world environments. The authors of \cite{Corridor} analyzed the impact of RIS installed at a corridor corner on THz channel using RT software. The simulation results demonstrated a significant reduction in path loss (approximately 20 dB). In \cite{TUBSRISSIMULATION}, the authors analyzed the signal enhancement of THz RIS in indoor industrial scenarios with the help of the Simulator for Mobile Networks (SiMoNe) from Technische Universit\"at Braunschweig (TUBS).

Multi-hop RIS further enhances coverage through the deployment of multiple RISs. In \cite{MULTIHOP1}, a hybrid beamforming architecture is introduced to improve coverage for multi-hop, multi-user RIS-supported wireless THz communication. The authors of \cite{MULTIHOP2} evaluate the impact of deploying multiple RISs and the number of signal hops on the overall achievable data rate in SISO systems. With the proper deployment of multiple RISs, communication data rates are expected to be significantly enhanced.

By integrating the antenna array–based RIS model into the high-performance RT platform (CloudRT, http://raytracer.cloud/) \cite{CloudRT1}\cite{CloudRT2}, this paper investigates the THz signal coverage enhancement effects of RIS in a hall scenario. For indoor regions with poor signal coverage, the effect of single-hop and dual-hop RIS deployments on received signal enhancement is analyzed in different deployment schemes. This work provides insights for the practical deployment of THz RIS and signal coverage estimation in real-world scenarios. 

The rest of the paper is structured as follows. Section II analyzes characteristics of signal coverage in different regions of the indoor hall scenario without RIS deployment.
Section III introduces the integration of the RIS model with the RT simulation platform. Section IV presents the coverage effects of single-hop RIS and dual-hop RIS in the indoor hall scenario. Section V concludes this paper.

\section{Power coverage of indoor hall scenario without RIS}

To simulate the channel characteristics of an indoor hall scenario, a three-dimensional (3D) hall model is first constructed on the SkecthUp software as shown in Fig. 1 (a). In this figure, the ceiling is hidden to provide a clearer view of the internal structure. The dimensions of the hall are 43 m $\times$ 45 m $\times$ 3 m. Fig. 1 (b) shows the internal view of the hall model. A frequency of 332 GHz is chosen for this study, and the electromagnetic parameters of the materials at this frequency are obtained from literature: the ceiling of the hall model is set to the Ceiling board \cite{MaterialITU}, the floor is set to the Granite \cite{Material}, the pillars and walls are set to the Concrete \cite{Material}. The transmitter (Tx) in this paper is consistently positioned at (5, 28, 1.5) m.

\begin{figure}
	\centering
	\includegraphics[width=0.9\columnwidth]{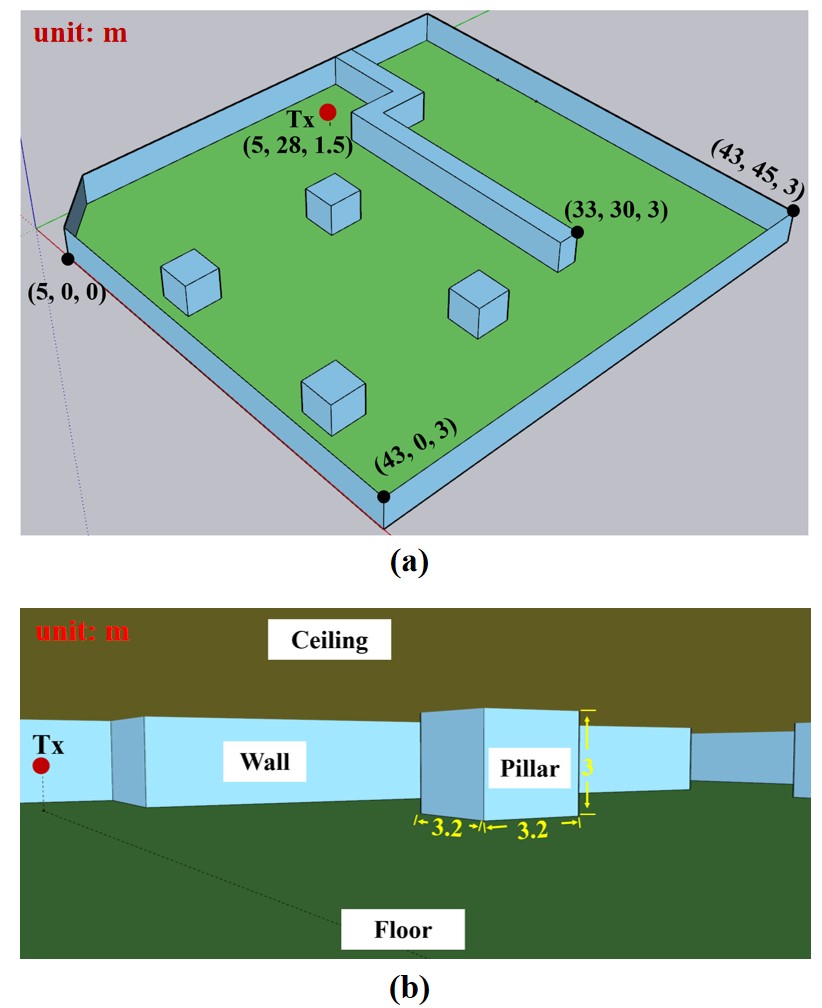}

	\caption{The (a) overall view and (b) internal view of the 3D hall.}
	\label{fig:eucap}
\end{figure}

To initially analyze the coverage performance of THz signals, both Tx and receivers (Rxs) antennas are configured as omni-directional. The  line-of-sight (LOS) path, up to the second-order reflected paths and first-order scattered paths are considered in the simulation. RT simulation is then performed, with the results shown in Fig. 2. It can be observed that in areas obstructed by pillars (such as Region A), the signal strength is relatively weaker compared to regions with LOS propagation paths. For areas obstructed by walls but close to wall openings (such as Region B), weak reflected and scattered paths are present. In areas obstructed by walls and far from wall openings (such as Region C), only sparse scattered paths or second-order reflected paths exist, resulting in extremely weak received signals. In some Rx positions (such as (15, 33, 1.5) m), no second-order reflected path or first-order scattered path are detected.
\begin{figure}
	\centering
	\includegraphics[width=0.8\columnwidth]{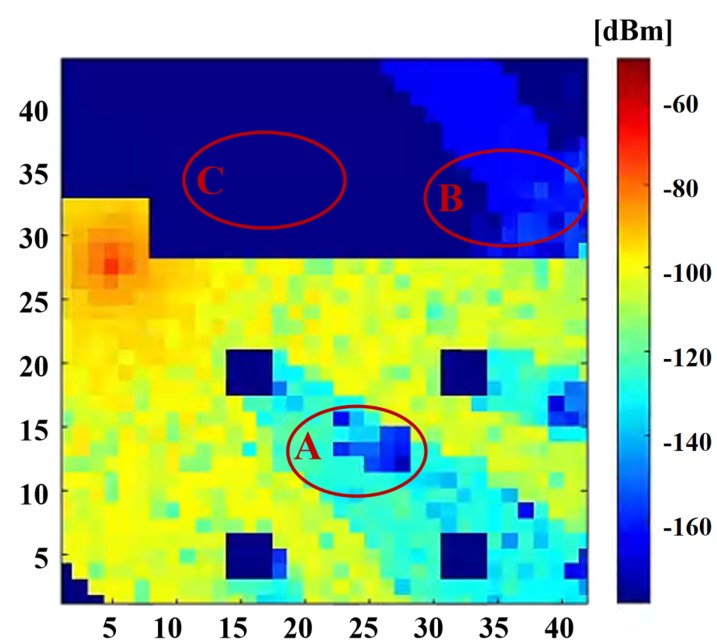}
	
	\caption{Heatmap of received power for the indoor hall scenario without RIS deployment.}
	\label{fig:eucap}
\end{figure}

\section{Integration of the RIS with the RT simulation platform}

The CloudRT simulation platform can effectively compute propagation mechanisms such as LOS, reflection, and scattering. Previous work has explored methods for integrating RIS and CloudRT, such as the approach proposed by the authors of \cite{Ailin}, which employs segmented channel simulations to achieve the integration of RIS into RT in tunnel scenarios. This paper is different from above approach in both the RIS radiation pattern and channel computations.

\subsection{RIS Radiation Pattern}
To effectively enhance the coverage performance in regions with weak signal strength, an antenna array-based RIS, which uses microstrip patches as the RIS elements, is incorporated into the RT simulation. The gain pattern of the RIS is calculated by extending the functions provided in MATLAB's array antenna toolbox \cite{TUBSRISSIMULATION}\cite{MATLAB}. In this paper, RIS achieves beam-steering by manipulating the phase of each RIS element. However, when the wave from Tx is obliquely incident on the RIS surface, the incident phase difference at each RIS element and the effective illuminated area must be considered. After considering the phase differences of the incident waves at each RIS element, the phase that needs to be applied by the RIS can be expressed as \cite{MATLAB}:
\begin{equation}\label{equ:pythagoras}
	{\footnotesize
	\begin{aligned}
		\varphi &= 2\pi \cdot 
		\frac{ \operatorname{round}\!\left(
			\operatorname{reminder}\!\left(
			-\tfrac{f}{c}\bigl(-\hat{r}_{\mathrm{beam}}\cdot\overrightarrow{OP} 
			+ \hat{r}_{\mathrm{i}}\cdot\overrightarrow{OP}\bigr),\,
			1
			\right) \cdot 2^N
			\right)}{2^N}
	\end{aligned}
}
\end{equation}

\begin{figure*}[!t]
	\centering
	\includegraphics[width=2\columnwidth]{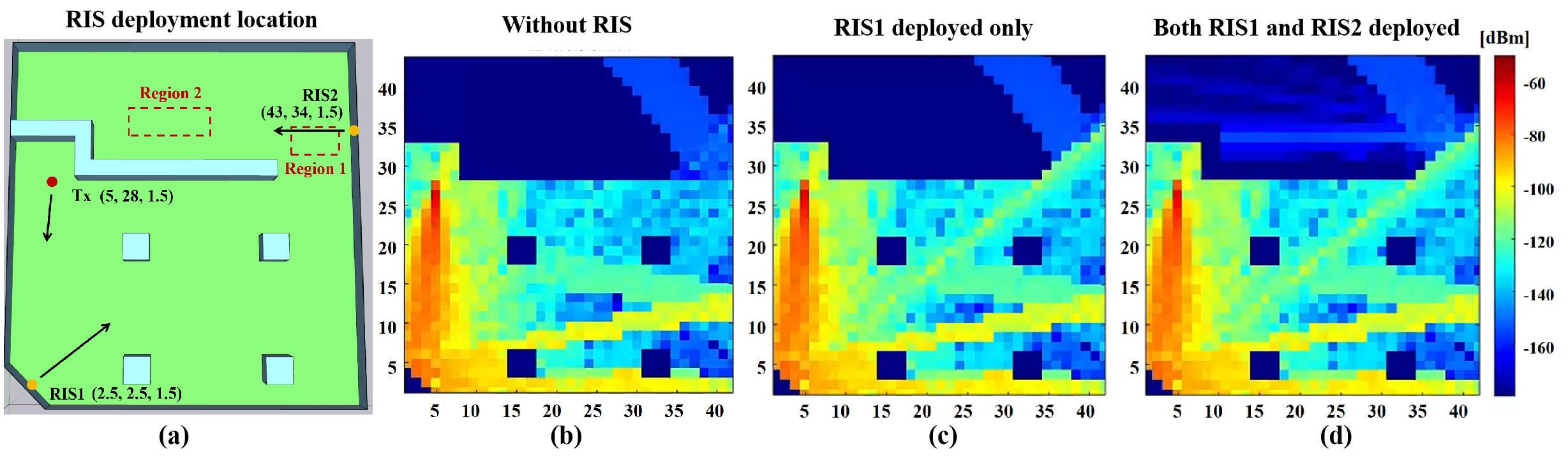}
	
	\caption{(a) RIS deployment scheme 1, where black arrows represent the radiation direction of the RIS or antenna, and red dashed boxes denote the regions to calculate average received power. Received power heatmaps are shown for (b) without RIS deployment, (c) only RIS1 deployed, and (d) both RIS1 and RIS2 deployed.}
	\label{fig:eucap}
\end{figure*}

\noindent where $\hat{r}_{\mathrm{i}}$ denotes the unit vectors of the incident direction, $\hat{r}_{\mathrm{beam}}$ is the unit vectors of radiation (main lobe) direction, $f$ is the frequency and $c$ denotes the speed of light. $\overrightarrow{OP}$ represents the vector pointing from the RIS centre point O to the RIS unit P. $N$ is the number of phase quantization bits ($N \geq 1$)
, which is an integer. The function round() is used to approximate to the nearest integer. The function reminder(a, b) returns the remainder when a is divided by b.

\subsection{RIS-assisted Channel}
To integrate the RIS model into this RT simulation platform, the following equation is used to calculate the received power of the single-hop RIS path \cite{channelmodel}: 
\begin{equation}
	{\footnotesize
	P = P_t \cdot G_{Tx} \cdot \frac{1}{4\pi L_{1}^2} \cdot \cos \theta_{1} \cdot S_{1} 
	\cdot F_{1} \cdot \frac{1}{4\pi L_{2}^2} \cdot \frac{G_{Rx}\lambda^2}{4\pi}
	}
\end{equation}
where $P_t$ is the transmitted power. $G_{Tx}$ and $G_{Rx}$ are the gains of the Tx antenna and Rx antenna, respectively. $L_{1}$ and $L_{2}$ are distances of RIS from Tx and Rx. $\lambda$ is the wavelength. $\cos\theta_1\cdot S_{1}$ is the effective incident area. $F_1$ is the radiation gain of RIS.

To enhance the signal enhancement effect of the RIS, the work \cite{danxinyi} suggests that RIS should be deployed in regions with a high density of first-order scattering points.
However, in Region C shown in Fig. 2, some Rx points may lack accessible first-order scattered paths, limiting the effectiveness of the single-hop RIS deployment. To address this issue, this work further explores the deployment of dual-hop RIS, with the following equation to calculate the received power \cite{channelmodel}:

\begin{equation}
		\begin{aligned}
			P =\ & P_t \cdot G_{Tx} \cdot \frac{1}{4\pi L_{1}^2}  
			\cdot \frac{\cos \theta_{1} \cdot S_{1} \cdot F_{1}}{4\pi L_{2}^2} \\
			& \cdot \frac{\cos \theta_{2} \cdot S_{2} \cdot F_{2}}{4\pi L_{3}^2} 
			\cdot \frac{G_{Rx}\lambda^2}{4\pi}
		\end{aligned}
\end{equation}
where $L_{1}$, $L_{2}$, $L_{3}$ are distances of Tx-RIS1, RIS1-RIS2, RIS2-Rx, respectively. $\cos\theta_1\cdot S_{1}$ and $F_1$ are the effective incident area and the radiation gain of RIS1. $\cos\theta_2\cdot S_{2}$ and $F_2$ are the effective incident area and the radiation gain of RIS2. RIS1 is the first-hop RIS, while RIS2 is the second-hop RIS.

\section{RIS-assisted signal coverage of indoor hall scenario}
In this section, three RIS deployment schemes are introduced to enhance the signal coverage in the indoor hall scenario presented in Section II. To enhance the radiation performance of the RIS, in the simulation setup, the Rxs remain the omni-directional antenna, while the Tx is set as a directional antenna with its maximum gain directed toward the position of RIS1. The transmitted power of the Tx is 0~dBm. The main lobe of the RIS1 radiation pattern is steered towards RIS2 to establish an efficient link. The number of RIS elements for both RIS1 and RIS2 is 100 $\times$ 100, with the element space of half a wavelength. Additionally, in the simulations conducted in this paper, only the waves directly radiated from the Tx to RIS1 are considered for the incident waves on RIS1. For waves incident on RIS2, only the waves directly radiated from RIS1 to RIS2 are taken into account. To simplify the analysis, it is assumed that the electromagnetic waves reflected by RIS1 and RIS2 are no longer reflected or scattered by other materials in the environment.

\subsection{RIS Deployment Scheme 1}

As shown in Fig. 3 (a), RIS1 and RIS2 are mounted on walls at different positions, located at (2.5, 2.5, 1.5) m and (43, 34, 1.5) m, respectively. Comparing Fig. 3 (b), (c) and (d), it can be observed that deploying RIS1 significantly enhances the received power in Region 1, while deploying both RIS1 and RIS2 simultaneously improves the received power in Region 2. Table I summarizes the average received power in two regions. For Region 1, deploying only RIS1 resulted in an average received power of -132.6 dBm, with an average received power gain of 20.3 dB compared to the result without RIS. For Region 2, deploying both RIS1 and RIS2 simultaneously achieved an average received power of -166.6 dBm. Since some locations in Region 2 lack second-order reflected paths and first-order scattered paths when there is no RIS deployment, the average received power is not calculated in Table I.

\vspace{-5pt}   
	
\begin{table}[h]
	\renewcommand{\arraystretch}{1.4}  
	\caption{The average received power in Region 1 and Region 2 for the RIS deployment scheme 1 }
	\label{table_example}
	\centering
	\begin{tabular}{c|c|c}
		\hline
		\noalign{\hrule height 1pt}  
		 & Region 1 & Region 2 \\
		\hline
		Without RIS & -152.9 dBm & - \\
		\hline
		RIS1 deployed only & -132.6 dBm & - \\
		\hline
		Both RIS1 and RIS2 deployed & -131.9 dBm & -166.6 dBm \\
		\hline
		\noalign{\hrule height 1pt}  
	\end{tabular}
\end{table}
\begin{figure*}[!t]
	\centering
	\includegraphics[width=2\columnwidth]{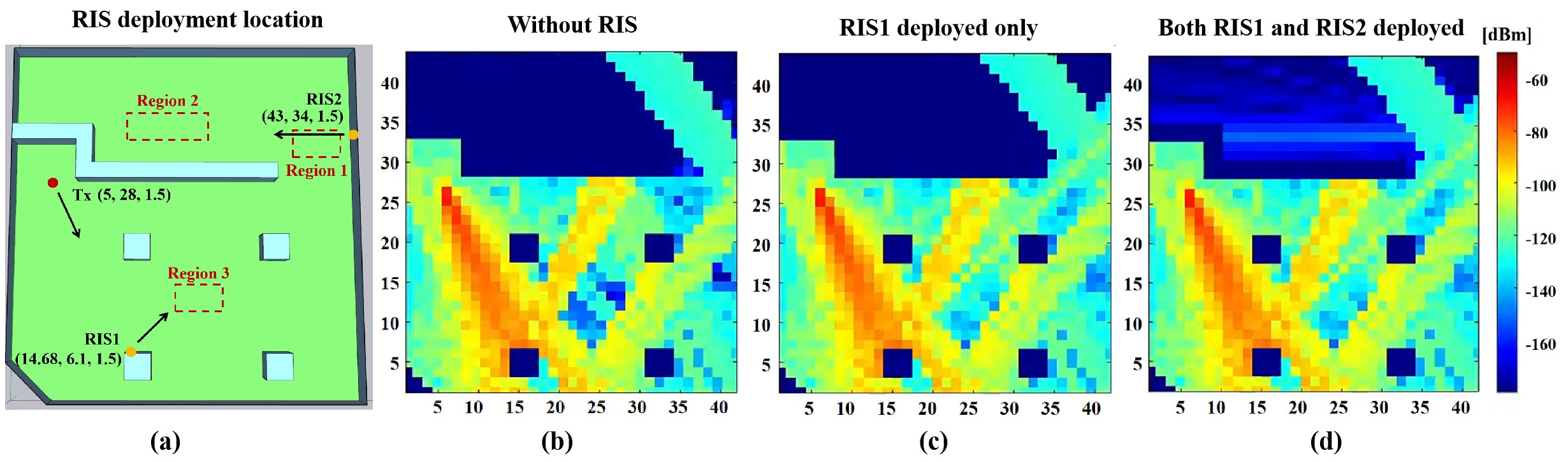}
	
	\caption{(a) RIS deployment scheme 2, where black arrows represent the radiation direction of the RIS or antenna, and red dashed boxes denote the regions to calculate average received power. Received power heatmaps are shown for (b) without RIS deployment, (c) only RIS1 deployed, and (d) both RIS1 and RIS2 deployed.}
	\label{fig:eucap}
\end{figure*}

\begin{figure*}[!t]
	\centering
	\includegraphics[width=2\columnwidth]{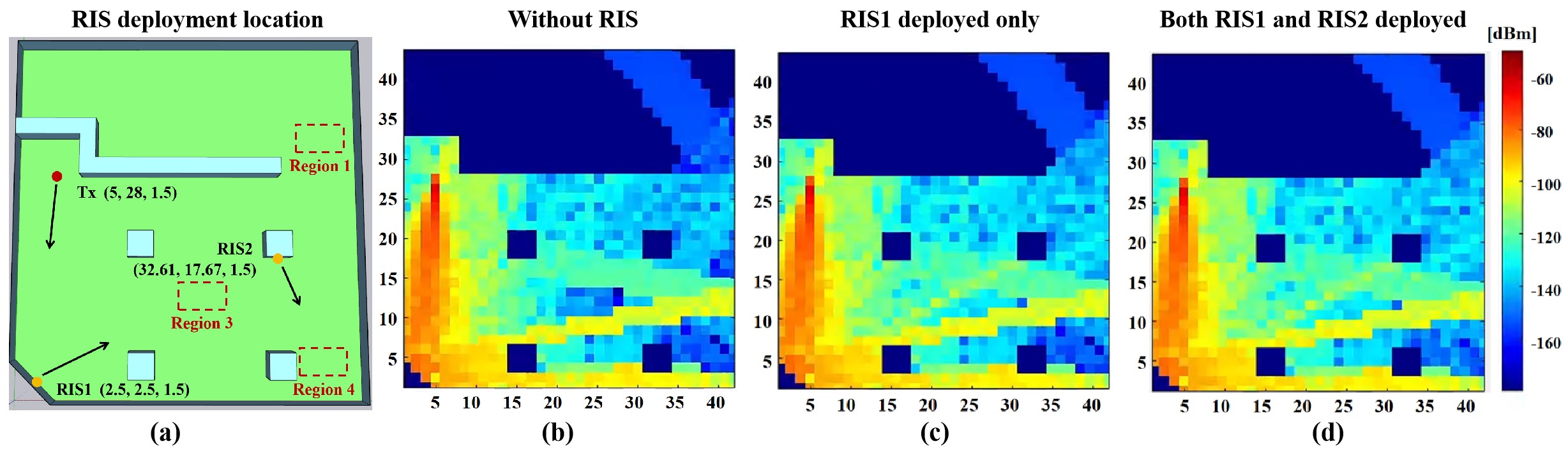}
	
	\caption{(a) RIS deployment scheme 3, where black arrows represent the radiation direction of the RIS or antenna, and red dashed boxes denote the regions to calculate average received power. Received power heatmaps are shown for (b) without RIS deployment, (c) only RIS1 deployed, and (d) both RIS1 and RIS2 deployed.}
	\label{fig:eucap}
\end{figure*}

\subsection{RIS Deployment Scheme 2}
As shown in Fig. 4 (a), RIS1 and RIS2 are mounted on walls at (14.68, 6.1, 1.5) m and (43, 34, 1.5) m, respectively. Due to the change of Tx orientation, the signal coverage has also been altered accordingly. 
Comparing Fig. 4 (b), (c) and (d), it can be observed that the received power in Region 1 and Region 3 is significantly enhanced with the deployment of RIS1, while received power enhancement is achieved in Region 2 with the simultaneous deployment of both RIS1 and RIS2. The average received power in three regions is summarized in Table II.
For Regions 1 and 3, compared to the result without RIS deployment, deploying only RIS1 achieves average received power gains of 7.2 dB and 10.0 dB, respectively. For Region 2, deploying both RIS1 and RIS2 simultaneously achieved an average received power of -162.2 dBm.

\begin{table}[h]
	\renewcommand{\arraystretch}{1.4}  
	\caption{The average received power in three regions for the RIS deployment scheme 2}
	\label{table_example}
	\centering
	\begin{tabular}{c|c|c|c}
		\hline
		\noalign{\hrule height 1pt}  
		& Region 1 & Region 2  & Region 3 \\
		\hline
		Without RIS & -128.3 dBm & - & -138.1 dBm \\
		\hline
		RIS1 deployed only & -121.1 dBm & - & -128.1 dBm \\
		\hline
	\makecell{Both RIS1\\and RIS2 deployed}
	 & -121.0 dBm & -162.2 dBm & -128.1 dBm \\
		\hline
		\noalign{\hrule height 1pt}  
	\end{tabular}
\end{table}

\subsection{RIS Deployment Scheme 3}

As shown in Fig. 5 (a), RIS1 and RIS2 are mounted on walls at (2.5, 2.5, 1.5) m and (32.61, 17.67, 1.5) m, respectively. RIS2 aims to enhance signal coverage in shadowed areas caused by obstructions of pillars. The average received power in three regions is summarized in Table III. Comparing Fig. 5 (b) and (c), it can be observed that deploying RIS1 significantly enhances power coverage in Region 1 and Region 3, with an average received power gain of 6.8 dB and 13.9~dB, respectively. The average received power in Region 4 increased by 4.9 dB because of the radiation from RIS2.

\begin{table}[h]
	\renewcommand{\arraystretch}{1.4}  
	\caption{The average received power in three regions for the RIS deployment scheme 3}
	\label{table_example}
	\centering
	\begin{tabular}{c|c|c|c}
		\hline
		\noalign{\hrule height 1pt}  
		& Region 1 & Region 3  & Region 4 \\
		\hline
		Without RIS & -152.9 dBm & -139.3 dBm & -150.8 dBm \\
		\hline
		RIS1 deployed only & -146.1 dBm & -125.4 dBm & -150.8 dBm \\
		\hline
		\makecell{Both RIS1\\and RIS2 deployed}
		& -146.1 dBm & -125.3 dBm & -145.9 dBm \\
		\hline
		\noalign{\hrule height 1pt}  
	\end{tabular}
\end{table}




\section{Conclusion}
Combining RT simulations, this paper analyzes the gain effects on indoor hall signal coverage achieved by single-hop and dual-hop RIS deployments under three different deployment schemes. 
In the case of the first-hop RIS, the gain is particularly significant in regions lacking the LOS path and where scattered and reflected paths are relatively weak, such as in Region 1 of the deployment scheme 1, the average received power achieved a gain of 20.3 dB. When two regions with poor signal coverage are aligned along the same line, it is feasible to enhance coverage using a single RIS, as demonstrated in Regions 1 and 3 of the deployment scheme 2, where average received power increased by 7.2 dB and 10 dB~respectively.
For the second-hop RIS, although its radiation power is relatively weaker, RIS2 can still enhance signal coverage in regions where there is no LOS path, and the scattered and reflected paths are extremely weak, such as Region 2 in the deployment schemes 1 and 2. 
In the future, we will investigate optimal deployment schemes for indoor multi-RIS collaboration to achieve the best coverage performance.


\section*{Acknowledgment}
This work is supported by Fundamental Research Funds for the Central Universities 2022JBQY004, Beijing Natural Science Foundation (L253023).
Dr. Qi Luo would like to acknowledge the support from TERRAMETA project, which is funded by the Smart Networks and Services Joint Undertaking (SNS JU) under the European Union's Horizon Europe research and innovation programme under Grant Agreement No 101097101, including top-up funding by UK Research and Innovation (UKRI) under the UK government's Horizon Europe funding guarantee.



%

\end{document}